\documentclass[reprint,aps,prc,amsmath,amssymb,nofootinbib,superscriptaddress]{revtex4-2}

\usepackage[T1]{fontenc}
\usepackage[utf8]{inputenc}
\usepackage{graphicx,color,rotating,pifont}
\usepackage{mathtools}
\usepackage{siunitx}
\usepackage{bm}
\usepackage{ae}

\usepackage{siunitx}
\usepackage{dcolumn}
\usepackage{txfonts}
\usepackage{tensor}
\usepackage{braket}
\usepackage{booktabs}
\usepackage{xspace}
\usepackage{mathrsfs}
\usepackage{amssymb} 
\usepackage{amsmath}
\usepackage{esvect}
\usepackage{diagbox}

\usepackage{hyperref}

\DeclareSIUnit{\fm}{\femto\meter}

\newcommand{\beq}{\begin{equation}}
\newcommand{\eeq}{\end{equation}}
\newcommand{\beqn}{\begin{eqnarray}}
\newcommand{\eeqn}{\end{eqnarray}}
\newcommand{\bsub}{\begin{subequations}}
\newcommand{\esub}{\end{subequations}}
\newcommand{\bpm}{\begin{pmatrix}}
\newcommand{\epm}{\end{pmatrix}}

\begin{document} 

\title{Emulating generator coordinate method with extended eigenvector continuation: Lipkin-Meshkov-Glick model}
 
\author{Q. Y. Luo} 
 \affiliation{School of Physics and Astronomy, Sun Yat-sen University, Zhuhai 519082, P.R. China} 

\author{X. Zhang} 
 \affiliation{School of Physics and Astronomy, Sun Yat-sen University, Zhuhai 519082, P.R. China} 

\author{L. H. Chen} 
 \affiliation{School of Physics and Astronomy, Sun Yat-sen University, Zhuhai 519082, P.R. China} 

\author{J. M. Yao} 
\email{Corresponding author: yaojm8@sysu.edu.cn}
\affiliation{School of Physics and Astronomy, Sun Yat-sen University, Zhuhai 519082, P.R. China} 

\date{\today}

\begin{abstract}
We present a benchmark study of generator coordinate method (GCM) combined with eigenvector continuation (EC) in two different schemes for the low-lying states of Lipkin-Meshkov-Glick (LMG) model, where the interaction strength is treated as a controlling parameter, simulating quantum  many-body systems with the phase transition from non-collective to collective states.  We demonstrate that the EC$_{\rm kmax}$ scheme accurately reproduces the low-lying states of the LMG model. In this scheme, the EC basis consists of the wave functions of low-lying states up to the $k_{\rm max}$-th state of sampling Hamiltonians. Compared to EC$_1$, which only includes the wave functions of the $k$-th state of sampling Hamiltonians for the $k$-th state of a target Hamiltonian, the EC$_{\rm kmax}$ scheme exhibits significantly improved efficiency and accuracy. This study suggests the potential utilization of the extended EC scheme as an efficient emulator for GCM calculations.

\end{abstract}

\maketitle

 \section{Introduction}

Generator coordinate method (GCM)~\cite{Hill:1953,Griffin:1957} is an important tool for modeling large-amplitude collective motions in atomic nuclei, including collective excitations\cite{Sheikh2021_JPG48-123001,Zhou:2023IJMPE}, dynamics of  clusters~\cite{Freer:2018RMP,Zhou:2019}, and nuclear fissions~\cite{Schunck:2016RPP,Verriere:2020FP,Bertsch:2022PRC,Li:2023PRC}. Recently, it has attracted increasing interest as it has been utilized to extend {\em ab initio} methods based on nuclear chiral interactions to study the low-lying states~\cite{Yao:2020PRL,Frosini:2021c} and giant monopole resonance \cite{Porro:2024_S1,Porro:2024tzt_S2} of medium-mass deformed nuclei, as well as to determine the nuclear matrix elements (NMEs) of candidates for neutrinoless double-beta ($0\nu\beta\beta$) decay~\cite{Yao:2022PPNP,Belley:2024}. 
The exact wave functions of nuclear low-lying states can, in principle, be well represented with the GCM ansatz if a sufficient number of generator coordinates are chosen. However, both complexity and computational time grow rapidly with the number of generator coordinates. Because of this complexity, quantifying the statistical uncertainty of GCM-based approaches for nuclear low-lying states has been a long-standing challenge. Therefore, there is considerable interest in finding an efficient optimization method or an emulator for GCM-based approaches.

Recently, statistical machine-learning techniques, combined with the subspace-selection algorithm based on orthogonality conditions~\cite{Romero:2021PRC}, were utilized to optimize GCM calculations for nuclear low-lying states and NMEs of $0\nu\beta\beta$ based on various nuclear Hamiltonians or energy density functionals (EDFs)~\cite{Zhang:2023PRC}. However, extending this optimization method to quantify uncertainties arising from the parameters of Hamiltonians or EDFs remains challenging, as it necessitates a significant number of repetitive GCM calculations with different parameter samplings. One potential solution to address this challenge is to develop an efficient emulator for GCM calculations.
 
In recent years, the eigenvector continuation (EC) method~\cite{Frame:2018PRL}, a special variant of reduced basis methods~\cite{Quarteroni:2013,Bonilla:2022PRC}, has emerged as a widely implemented technique for emulating few- and many-body calculations. The basic idea of the EC method is representing the eigenvector of a target Hamiltonian within a low-dimensional manifold formed by the eigenvectors (also known as training vectors) of a set of sampling Hamiltonians. The smoother the manifold, the fewer training vectors are needed. The efficiency of the EC method in conjunction with other many-body methods has been demonstrated in various toy models~\cite{Franzke:2022PLB,Sowinski:2022PRC,Franzke:2024PRC} and  in the application to nuclear structure and scattering processes~\cite{Ekstrom:2019PRL,Konig:2020fn,Furnstahl:2020PLB,Drischler:2021PLB,Bai:2021PRC,Sarkar:2021}. For further insights into the EC method, readers are encouraged to refer to recent reviews~\cite{Drischler:2022,Duguet:2023RMP}

It is worth noting that the majority of applications of the EC method in nuclear structure are focused on the ground states of nuclei, whose wave functions are typically constructed as a linear combination of the ground-state wave functions of sampling Hamiltonians. There are only a few applications to excited states. Recently, the excited states of a harmonic oscillator were studied using EC combined with many-body perturbation theory (MBPT)~\cite{Franzke:2022PLB}, where the wave function of the $k$-th excited state of a target Hamiltonian is expanded in terms of the wave functions of the $k$-th states of the sampling Hamiltonians determined by the MBPT. This scheme is called EC$_1$ hereafter. In contrast, in the EC plus interacting shell model (ISM)~\cite{Yoshida:2022PETP}, the wave functions of the lowest $k_{\rm max}$ states of sampling Hamiltonians were included into the EC basis to expand  those of target Hamiltonians. We call this extended EC scheme  as EC$_{\rm kmax}$.  In the EC$_{\rm kmax}$ scheme, the EC basis form a complete basis as $k_{\rm max}\to \infty$, irrespective of the number of sampling Hamiltonians. However, in practical application, the number of the lowest states $k_{\rm max}$ is truncated to a finite number that gives a rather convergent solution. It has been shown in the EC$_{\rm kmax}$+ISM that the choice of $k_{\rm max}=5$ improves the relative error of the five lowest states of four $sd$-shell nuclei by a factor of two~\cite{Yoshida:2022PETP}. In this work, we examine the performance of these two EC schemes in the GCM calculations for the low-lying states of the Lipkin-Meshkov-Glick (LMG) model~\cite{Lipkin:1965,Meshkov:1965,Glick:1965}, and compare these results against the exact solutions of the diagonalization method. The LMG model is an exactly solvable model which has been widely used for testing various many-body approaches, including random-phase-approximation (RPA)\cite{Hagino:2000PRC} and GCM\cite{Severyukhin:2006}, as well as quantum computing algorithms~\cite{Romero:2022PRC,Hlatshwayo:2022,Beaujeault-Taudiere:2024PRC,Aggarwal:2024}.

The article is structured as follows. In Sec.\ref{sec:framework}, we present the main formulas of the LMG model, including the exact solution of the diagonalization method, Hartree-Fock (HF) approximation, GCM, and EC+GCM solutions. The results of calculations with different many-body methods are compared in Sec.\ref{sec:results}. Finally, a brief summary and outlook are provided in Sec.\ref{sec:summary}.
 
 \section{The Lipkin-Meshkov-Glick  model} 
\label{sec:framework}

\subsection{The Hamiltonian}

\begin{figure}[]
    \centering
    \includegraphics[width=6cm]{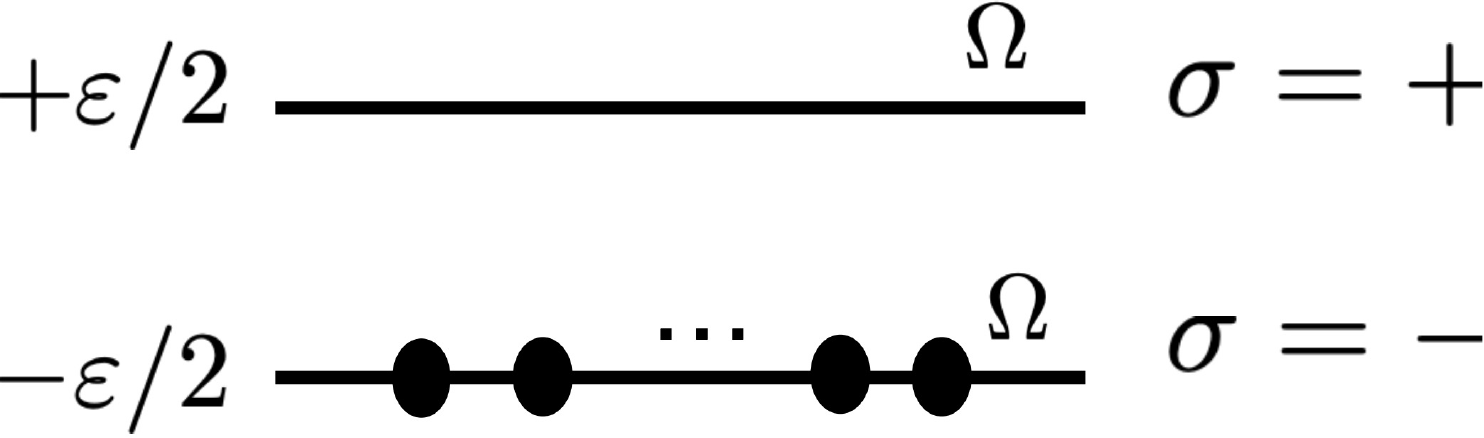}
    \caption{Illustration of the LMG model, where $N(=\Omega)$ identical fermions are distributed across two levels, each with a degeneracy of $\Omega$.}
    \label{fig:levels}
\end{figure}

The LMG model describes a system of $N(=\Omega)$ identical fermions distributed in two $\Omega$-fold degenerate levels labeled with $\sigma=\pm$, respectively. The energy gap between these two levels is $\varepsilon$. The LMG model is schematically depicted in Fig.\ref{fig:levels}. The Hamiltonian consists of a one-body term and a monopole-monopole two-body interaction term\cite{Lipkin:1965}
\beqn
\label{eq:Hamiltonian}
    \hat H = \varepsilon \hat K_0 - \frac12 V(\hat K_+\hat K_+ + \hat K_-\hat K_-)  
\eeqn
where $V$ is the two-body interaction strength and
\bsub\begin{align}
    \hat K_0 &= \frac12\sum^\Omega_{m=1}(c^\dagger_{+m}c_{+m} - c^\dagger_{-m}c_{-m}),\\
   \hat  K_+ &= \sum^\Omega_{m=1}c^\dagger_{+m}c_{-m}, \\
   \hat  K_- &= (\hat K_+)^\dagger.
\end{align}
\esub

The operators $c^\dagger_{+m}$ and $c^\dagger_{-m}$ create particles in the upper and lower levels, respectively. It can be proven that the operators $\hat K_0$, $\hat K_{\pm}$ satisfy the following commutation relations of angular momentum operators
\beqn 
[\hat K_0, \hat K_\pm]=\pm \hat K_\pm,\quad [\hat K_+, \hat K_-]=2\hat K_0.
\eeqn 
 
\subsection{Exact solution with the diagonalization method}

\begin{figure}[]
    \centering
    \includegraphics[width=\columnwidth]{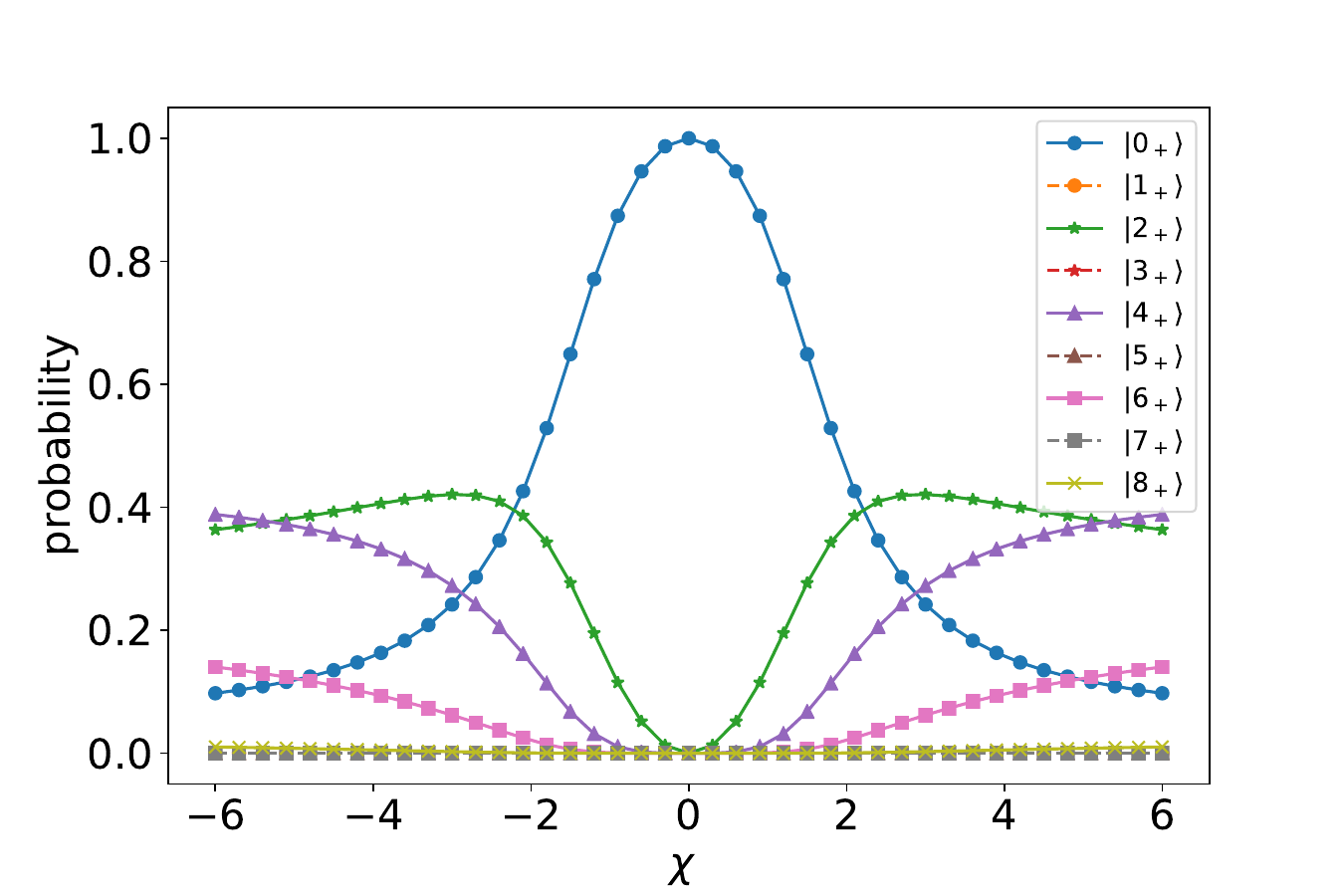}
    \caption{Probability distribution $|f^{k}{N+}|^2$ of the exact ground state of the Hamiltonian $\hat H(\chi)$ with different interaction parameters $\chi$, calculated in the basis $\ket{N{+}}$ as defined in Eq. (\ref{eq:basis_exact}), where $\Omega=8$.}
    \label{fig:exact_gs_wfs}
\end{figure}

The wave function for the $N$ particles in the LMG model can be expanded in the configurations basis
\beq 
\label{eq:wf_exact}
\ket{\Psi^k} = \sum_{N_+} f^k_{N_+} \ket{N, N_{+}},
\eeq 
where the superscript $k$ distinguishes different states, and $N_{+}$ represents the number of particle-hole ($ph$) excitations, i.e., the number of particles excited from the lower energy level $(\sigma=-)$ to the  upper energy level $(\sigma=+)$, with its value $N_+\in [0, N]$. If one introduces quasispin $J=N/2$ and its projection $M=N_+-N/2$, then the basis $\ket{N, N_{+}}$  can be rewritten as
\beq 
\label{eq:basis_exact}
\ket{N, N_{+}} \equiv \ket{N/2, N_+-N/2} = \ket{J,M},
\eeq 
with $M=-J,\cdots,J-1, J$. The dimension of the basis is $2J+1=N+1$. The operators $\hat K_0, \hat K_{\pm}$ are then interpreted as quasispin operators with the following relations~\cite{Varshalovich:1988}
\bsub\beqn
\hat K_0 \ket {J,M} &=& M \ket {J,M},\\
\hat{K}_{\pm}\ket {J,M} &=&  \sqrt{(J\mp M)(J\pm M+1)}\left|J,M\pm1\right\rangle, 
\eeqn
\esub
from which one finds the  expression for the matrix elements of the Hamiltonian in the configuration basis,  
\beqn
\label{eq:basis}
\left\langle N, N_{+}^{\prime}|\hat{H}(\chi)| N, N_{+}\right\rangle
&= & \varepsilon M
-\frac{V}{2}\Bigg[C_{+}(M) C_{+}(M+1) \delta_{N_{+}^{\prime}, N_{+}+2}  \nonumber\\
&&+C_{-}(M) C_{-}(M-1) \delta_{N_{+}^{\prime}, N_{+}-2}\Bigg]
\eeqn 
with $C_{ \pm}(M) =\sqrt{J(J+1)-M(M \pm 1)}$. One can observe that the matrix elements of the Hamiltonian are zero if $N_+$ and $N'_+$ have opposite number parity. This implies that the space formed by $\ket{N, N_+}$ can be divided into two subspaces with even and odd-number parity, respectively. The energy $E^k$ and expansion coefficient $f^k_{N_+}$ of the wave function (\ref{eq:wf_exact}) for the $k$-th state are obtained from the diagonalization of the Hamiltonian matrix $\left\langle N, N_{+}^{\prime}|\hat{H}(\chi)| N, N_{+}\right\rangle$. For a small value of $\Omega$, it is not difficult to derive analytical solutions~\cite{Lipkin:1965}. For instance, there are 9 solutions for $\Omega=8$,\footnote{In the original paper~\cite{Lipkin:1965}, a factor of 6 is missing in the last four solutions.}
\beqn 
\frac{E}{\varepsilon}
&=&0, \pm\Bigg[5+\frac{113}{7^2}\chi^{2} \pm 4\left(1+\frac{38}{7^2}\chi^{2}+\frac{550}{7^4}\chi^{4}\right)^{\frac{1}{2}}\Bigg]^{\frac{1}{2}},\nonumber\\
&& \pm\Bigg[10+\frac{118}{7^2}\chi^{2} \pm6\left(1-\frac{2}{7^2}\chi^{2}+\frac{225}{7^4}\chi^{4}\right)^{\frac{1}{2}}\Bigg]^{\frac{1}{2}}, 
\eeqn 
where the interaction parameter $\chi$ is defined as follows,
\begin{equation}
    \chi = \frac{V}{\varepsilon}(\Omega-1).
\end{equation}

Figure~\ref{fig:exact_gs_wfs} illustrates the change in the probability of each basis state $\ket{N, N_{+}}$ in the exact ground state of the Hamiltonian $\hat{H}(\chi)$ as a function of the interaction parameter $\chi$ for the $\Omega=8$ case. As pointed out in Ref.\cite{Severyukhin:2006}, in the limit of $\chi \to0$, the $k$-th state will be the pure $\ket{N, N_+=k}$ component, corresponding to $kp$-$kh$ excitations. As the interaction strength $|\chi|$ increases, whether attractive or repulsive, each state becomes a complicated mixing of many $ph$ excitations. It is observed in Fig.\ref{fig:exact_gs_wfs} that the weights of components with more particles excited from the lower level to the upper level gradually increase. When $\chi$ increases beyond a critical value, the system undergoes a phase transition from {\em spherical} (or shell-model like) to {\em deformed} (collective) states. Notably, the probability distribution is symmetric with respect to $\chi=0$.

\subsection{HF solutions}
\begin{figure}[] 
    \includegraphics[width=0.95\columnwidth]{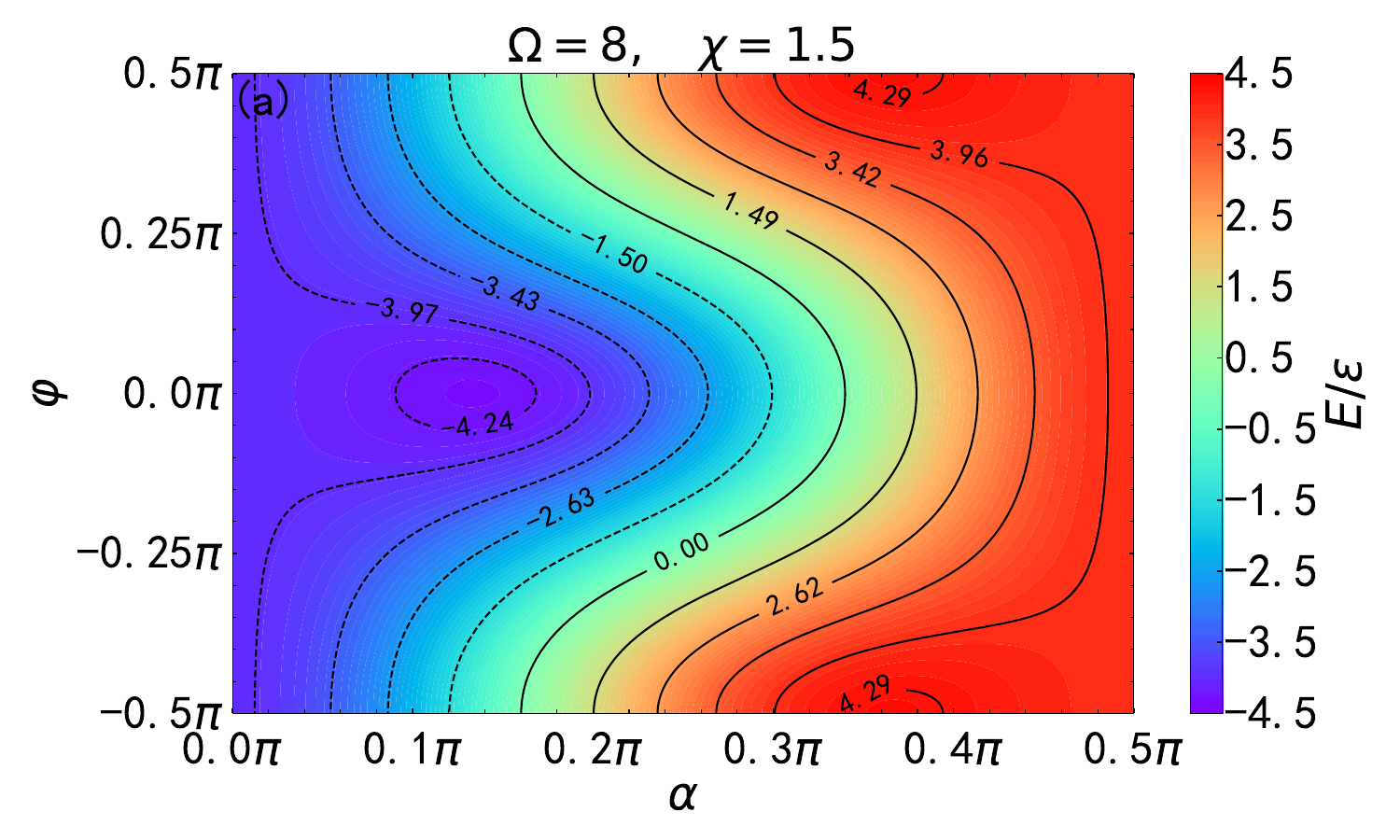} 
    \includegraphics[width=0.95\columnwidth]{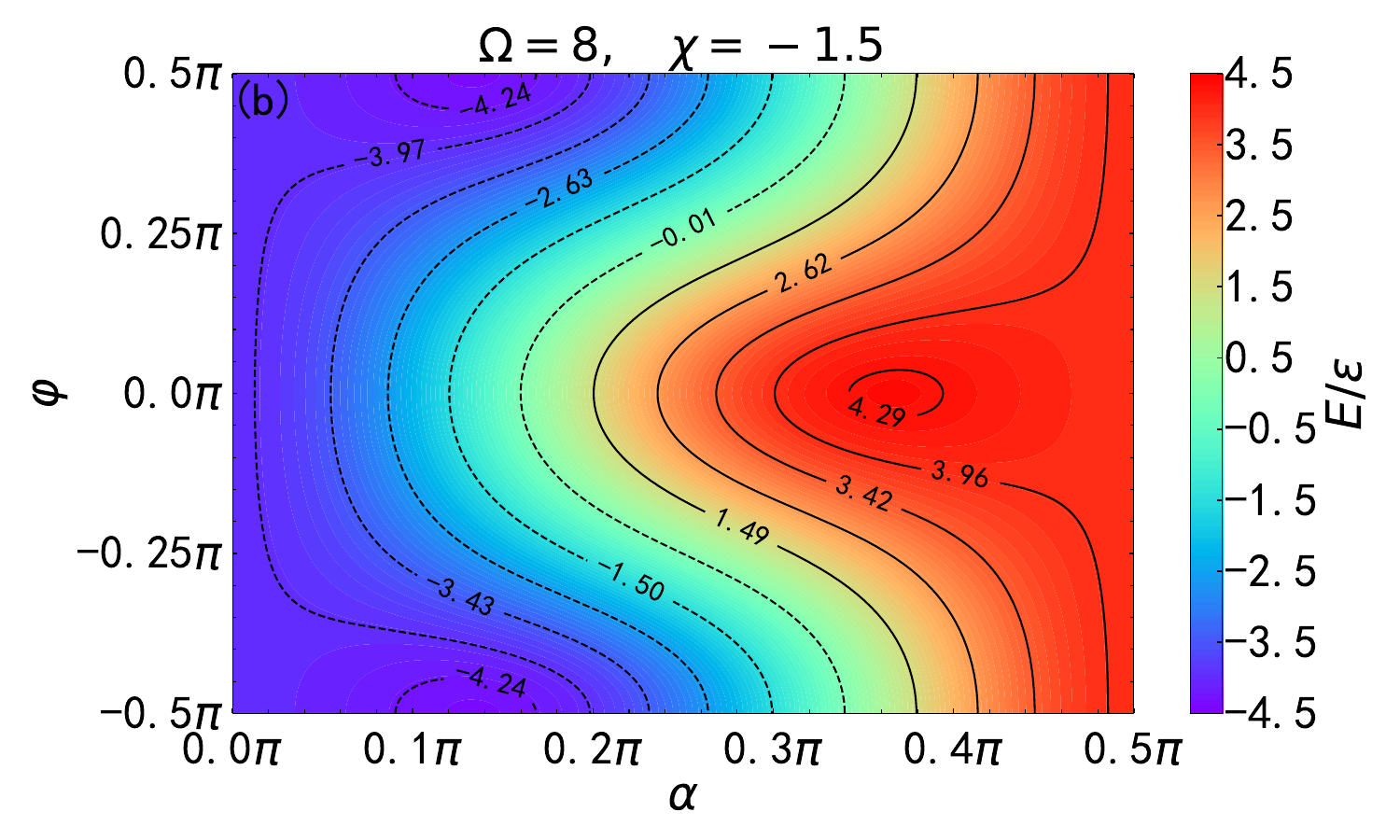} 
    \caption{Contour plots of the normalized energies $E_{\rm HF}/\varepsilon$ of HF states in the $(\alpha, \varphi)$ plane, where the interaction strength is chosen as $\chi=\pm1.5$, respectively.  }
    \label{fig:HF_PES}
\end{figure}

\begin{figure}[] 
    \includegraphics[width=0.95\columnwidth]{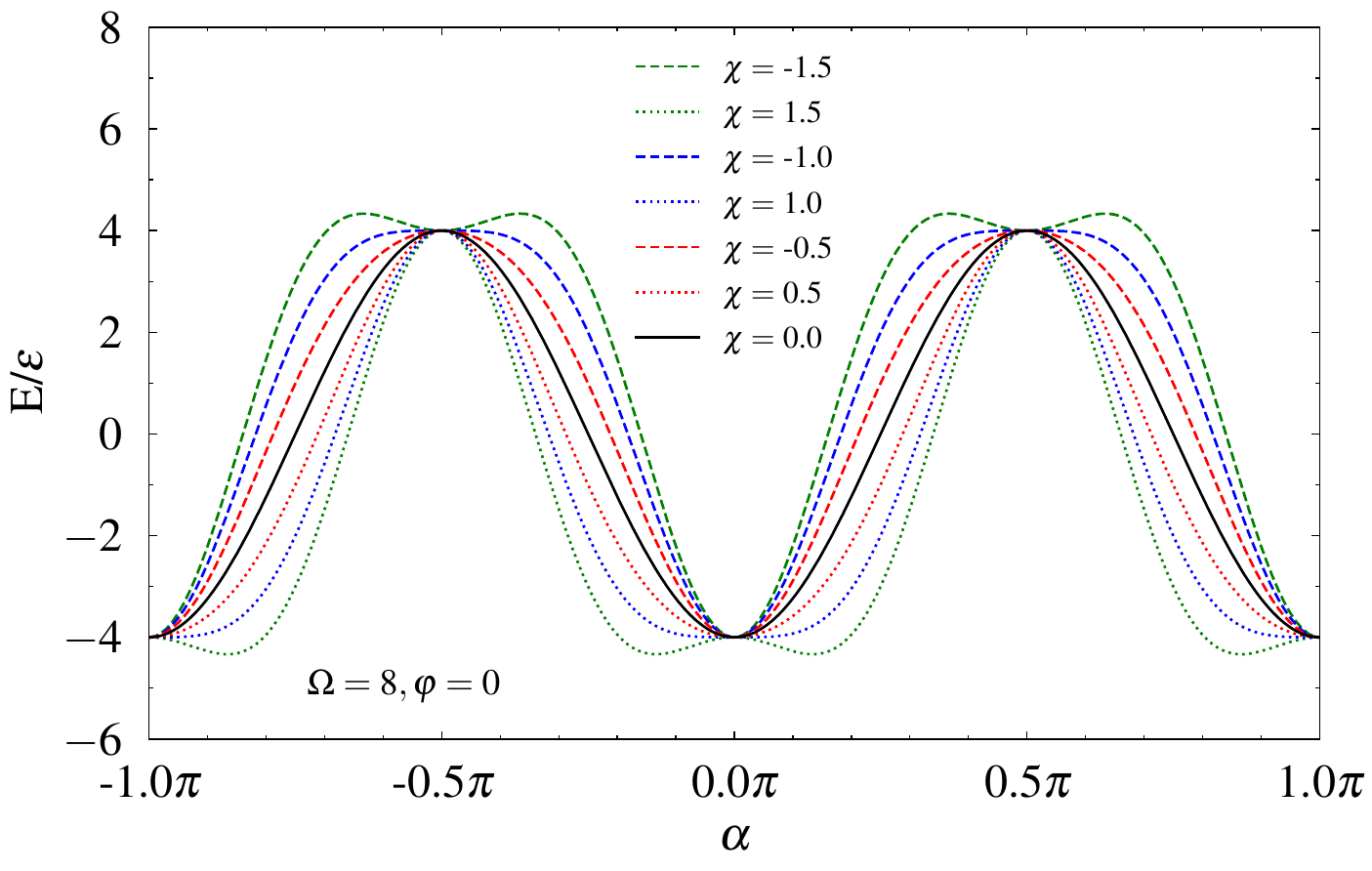}  
    \caption{The energies $E_{\rm HF}/\varepsilon$ of the HF states for the system with $\Omega=8$ as a function of the parameter $\alpha$, where  $\varphi=0$ and the interaction strength $\chi$ varies from $-1.5$ to $+1.5$.  }
    \label{fig:HF_Energy}
\end{figure}

In the HF approach, the ground-state wave function is approximated with a Slater determinant
\begin{align}
    \ket{\Phi(\alpha,\varphi)}= \prod^\Omega_{m=1}a^\dagger_{0m}(\alpha, \varphi)|-\rangle
\end{align}
where the particle creation operator $a^\dagger_{0m}(\alpha, \varphi)$ in the HF basis is related to the creation operator $c^\dagger_{\pm m}$ in the single-particle basis by~\cite{Severyukhin:2006}
\begin{equation}
\label{eq:particle_operators}
    \begin{pmatrix}
        a^\dagger_{0m}(\alpha,\varphi)\\
        a^\dagger_{1m}(\alpha,\varphi)
    \end{pmatrix} 
    = \begin{pmatrix}
        \cos\alpha&\sin\alpha e^{-i\varphi}\\
        -\sin\alpha e^{i\varphi}&\cos\alpha
    \end{pmatrix}
    \begin{pmatrix}
        c^\dagger_{-m}\\
        c^\dagger_{+m}
    \end{pmatrix}.
\end{equation}
The indices $0, 1$ denote hole and particle states, respectively. The two parameters $(\alpha, \varphi)$ are used to distinguish different HF states.  The expectation value of the Hamiltonian (\ref{eq:Hamiltonian}) with respect to this HF state can be derived analytically~\cite{Ring:1980},
\begin{equation}
\label{eq:Energy_HF}
    E_{\rm HF} = -\frac{\varepsilon}{2}\Omega\Big(\cos2\alpha+\frac12\chi\sin^22\alpha\cdot\cos2\varphi\Big),
\end{equation}
which for $\Omega=8$ and $\chi=\pm1.5$ is plotted in Fig.~\ref{fig:HF_PES}. 
It is observed that the energy minimum is located at $\alpha=\frac{1}{2}\cos^{-1}(1/|\chi|)=0.134\pi$, with $\varphi=0, \pm\pi/2$ for $\chi=1.5$ and $-1.5$, respectively. Figure~\ref{fig:HF_Energy} illustrates the normalized energies $E_{\rm HF}/\varepsilon$ of the HF states by the Hamiltonians with different values of $\chi$ as a function of the parameter $\alpha$, where $\varphi=0$.  According to Eq.(\ref{eq:Energy_HF}), the energy will be independent of the value of $\chi$ for $\varphi=\pi/4$. In particular, the energy curves with positive and negative values of $\chi$ are exchanged if the value of $\varphi$ is switched from $0$, to $\pm\pi/2$. It means that the value of $\varphi$ also controls the two-body correlation of the system. Notably, it is shown in Fig.~\ref{fig:HF_Energy} that the system undergoes phase transitions as the interaction parameter $\chi$ varies from $-1.5$ to $1.5$. For $|\chi|<1$, the energy minimum is located at the noncollective {\em spherical} shape with $\alpha=n\pi$. When $\chi>1$, the energy minimum is found at the collective {\em deformed} shape with $\alpha=\pm\frac{1}{2}\cos^{-1}(1/\chi)+n\pi$ and $\varphi=0$. Conversely, for $\chi<-1$, the energy minimum is located at $\alpha=\pm\frac{1}{2}\cos^{-1}(-1/\chi)+n\pi$ and $\varphi=\pm\pi/2+n\pi$, where $n$ is an integer. Since both the energies and wave functions are periodic functions of the parameters $\alpha$ and $\varphi$, we need only consider the HF states with these two parameters in the interval $[-\pi/2, \pi/2]$, subsequently.

\subsection{GCM solutions}

In the GCM, the wave function $\ket{\Psi^\kappa_{\rm GCM}(\chi)}$ is constructed as a linear combination of HF states $\ket{\Phi(\alpha,\varphi)}$ with different values of parameters $(\alpha,\varphi)$.  For the sake of simplicity, we introduce a symbol $\mathbf{q}$ to stand for $(\alpha,\varphi)$. In practical applications, the generator coordinates $\mathbf{q}$ are discretized.  The wave function $\ket{\Psi^\kappa}$ becomes 
\begin{equation}
\label{eq:GCM_wf}
    \ket{\Psi^\kappa_{\rm GCM}(\chi)} 
    = \sum^{N_{\mathbf{q}}}_{\mathbf{q}=1} f^\kappa(\chi; \mathbf{q})\ket{\Phi(\mathbf{q})},
\end{equation}
where $N_q$ represents the number of HF states which are independent of the interaction parameter $\chi$. Since the operator $a^\dagger_{0m}$ defined in (\ref{eq:particle_operators}) mixes the operators of $c^\dagger_{+m}$ and $c^\dagger_{-m}$, the HF state $\ket{\Phi(\alpha,\varphi)}$ does not have a definite number of particles in the upper or lower levels, even though the total particle number is still conserved.  Therefore, the non-orthogonal basis formed by $\{\ket{\Phi(\mathbf{q})}\}$ cannot be divided into two subspaces anymore.
The weight function $f^\kappa(\chi; \mathbf{q})$ is determined by the Hill-Wheeler-Griffin (HWG) equation,
\beq 
\sum^{N_{\mathbf{q}}}_{\mathbf{q}^{\prime}=1}\langle\Phi(\mathbf{q})|\hat H(\chi)|\Phi(\mathbf{q}^{\prime})\rangle 
- E^\kappa\langle\Phi(\mathbf{q})|\Phi(\mathbf{q}^{\prime})\rangle\Big) 
   f^\kappa(\chi; \mathbf{q}^{\prime}) = 0.
\eeq 
The norm kernel and Hamiltonian kernel of the GCM can be derived analytically~\cite{Severyukhin:2006}, 
 \beqn   
\label{eq:overlap}
&&\langle\Phi(\mathbf{q})|\Phi(\mathbf{q}^{\prime})\rangle =\Big(\mathcal{N}_{\mathbf{q}\mathbf{q}^{\prime}}\Big)^\Omega,\\ 
&&\langle\Phi(\mathbf{q})|\hat H(\chi)|\Phi(\mathbf{q}^{\prime})\rangle= -\frac{\varepsilon\Omega}{2} \notag\\ 
    &&\times \bigg\{\cos^2(\alpha)\cos^2(\alpha^{\prime}) 
    - \sin^2(\alpha)\sin^2(\alpha^{\prime})e^{2i(\varphi-\varphi^{\prime})}\notag\\
    && +\chi\Big[\sin^2(\alpha)\cos^2(\alpha^{\prime})e^{2i\varphi}  
    + \sin^2(\alpha^{\prime})\cos^2(\alpha) e^{-2i\varphi^{\prime}}\Big]\bigg\}
    \Big(\mathcal N_{\mathbf{q}\mathbf{q}^{\prime}}\Big)^{\Omega-2},\nonumber \\
\eeqn  
with  ${\cal N}_{\mathbf{q}\mathbf{q}^{\prime}}\equiv\cos(\alpha)\cos(\alpha^{\prime})+\sin(\alpha)\sin(\alpha^{\prime})e^{i(\varphi-\varphi^{\prime})}$.

\subsection{Emulating GCM solutions with the EC method}
In the EC+GCM, the wave function $\ket{\Psi^k(\chi_\odot)}$ of the $k$-th state for a target Hamiltonian $\hat H(\chi_\odot)$ is represented in a manifold ${\cal M}^{\rm EC}$ of the many-body Hilbert space, formed by the $N_{\rm EC}$ EC basis functions.
\begin{equation}
\label{eq:EC_GCM_wfs}
   \ket{\Psi^k_{\rm EC}(\chi_\odot)} =\sum^{k_{\rm max}\ge k}_{\kappa=1}\sum^{N_t}_{t=1}g^{k}(\kappa,\chi_t)\ket{\Psi^{\kappa}_{\rm GCM}(\chi_t)},
\end{equation}
where $N_t$ represents the number of sampling (training) Hamiltonians $\hat H(\chi_t)$ used to produce the set of EC basis functions $\ket{\Psi^\kappa_{\rm GCM}(\chi_t)}$ with $\kappa=1,2,\cdots, k_{\rm max}$. Thus, the dimension of the EC basis is $N_{\rm EC}=N_tk_{\rm max}$. This scheme is called EC$_{\rm kmax}$+GCM($N_t$), for convenience.  The weight  $g^k(\kappa,\chi_t)$ for the $k$-th state of the target Hamiltonian $\hat H(\chi_{\odot})$ is determined by the following generalized eigenvalue equation,
\beq 
 \sum^{k_{\rm max}}_{\kappa'=1}\sum^{N_{t}}_{t'=1} \Bigg[ 
 {\cal H}^{\kappa\kappa'}_{tt'}(\chi_\odot) 
 - E^k_{\chi_\odot}   {\cal N}^{\kappa\kappa'}_{tt'} \Bigg] g^k(\kappa',\chi_{t'})=0,
\eeq 
where the norm and Hamiltonian kernels of the EC method are defined  as
\bsub
\label{eq:EC_kernel}
\beqn 
     {\cal N}^{\kappa\kappa'}_{tt'} &=& \bra{\Psi^\kappa_{\rm GCM}(\chi_t)}\Psi^{\kappa'}_{\rm GCM}(\chi_{t'})\rangle,\\
    {\cal H}^{\kappa\kappa'}_{tt'}(\chi_\odot) &=& \bra{\Psi^\kappa_{\rm GCM}(\chi_t)}\hat H(\chi_\odot)\ket{\Psi^{\kappa'}_{\rm GCM}(\chi_{t'})}.
\eeqn
\esub

\subsection{Time complexities of GCM and EC+GCM}

Let's compare the time complexity of the GCM and EC+GCM methods for the low-lying states of $N_{\chi_\odot}$ target Hamiltonians, where $N_{\chi_\odot}$ could be on the order of $10^6$ for chiral Hamiltonians~\cite{Ekstrom:2019PRL}. The time complexity of the $N_{\chi_\odot}$ repetitive GCM calculations is
\begin{align}
\label{eq:time_complexity_GCM}
    T_{\rm GCM} =O\Big(N^2_qN_{\chi_\odot}\Big)\Delta T_1,
\end{align}
where  $\Delta T_1$ represents the time cost for computing each overlap (\ref{eq:overlap}) between two different HF states. For the EC+GCM($N_t$) with $N_t$ sampling Hamiltonians, one needs to explicitly evaluate all the EC kernels (\ref{eq:EC_kernel}). Thus, the time complexity is composed of two parts,
\begin{align}
\label{eq:time_complexity_EC}
    T_{\rm EC+GCM} = O\Big( N^2_q N^2_t \Big)\Delta T_1 + O\Big(N^2_{\rm EC}N_{\chi_\odot}\Big)\Delta T_2,  
\end{align}
where $\Delta T_2$ is the time cost for computing each  EC kernel. 
In the LMG model, since the HF states are independent of the interaction parameter $\chi$, the first term in (\ref{eq:time_complexity_EC}) simplifies to $O\Big( N^2_q N_t\Big)\Delta T_1$, representing $N_t$ times of computations of $N^2_q$ overlaps. Moreover, the computation time $\Delta T_1$ is negligible as one has an analytical expression for the overlaps. Therefore, the advantage of using EC+GCM for the LMG model is not obvious, if $\Delta T_2 \simeq \Delta T_1$. However, in many GCM studies based on EDFs or Hamiltonians defined in the full single-particle space~\cite{Sheikh2021_JPG48-123001,Yao:2022PPNP}, the most time-consuming part is the calculation of the overlaps, i.e., $\Delta T_1>>\Delta T_2$. In this case, it is expected that $T_{\rm EC+GCM}<<T_{\rm GCM}$ for $N_{\chi_\odot}\simeq 10^6$. In particular, the computation time can be significantly reduced if one can reduce the number of sampling Hamiltonians $N_t$. As it will be demonstrated, this could be achieved in the EC$_{\rm kmax}$ scheme.

 \section{Results and discussion}
 \label{sec:results}
 
\subsection{Results of GCM calculations}

\begin{figure}
    \centering
    \includegraphics[width=\columnwidth]{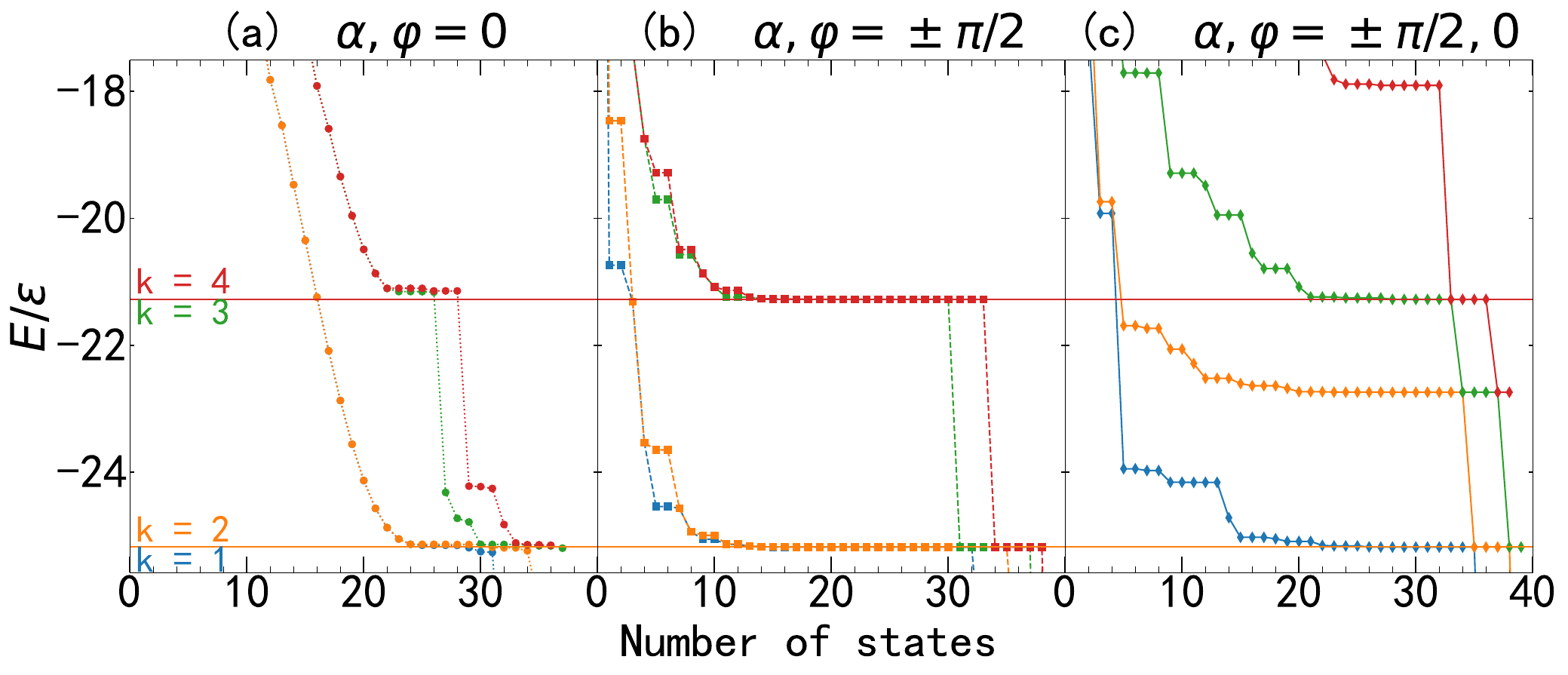}
    \caption{The convergence of state energies with the number of natural states for $\Omega=30$ particles in the LMG model with the interaction strength parameter $\chi=-3$. The first generator coordinate $\alpha$ ranges from $-\pi/2$ to $\pi/2$, with 40 equally distributed mesh points. The second generator coordinate $\varphi$ is fixed at (a) 0, (b) $\pm\pi/2$, and (c) $(0, \pm\pi/2)$, respectively.
The exact energies are indicated with horizontal lines.}
    \label{fig:E_plateau_GCM}
\end{figure}

\begin{figure}[]
    \centering
    \includegraphics[width=0.7\columnwidth]{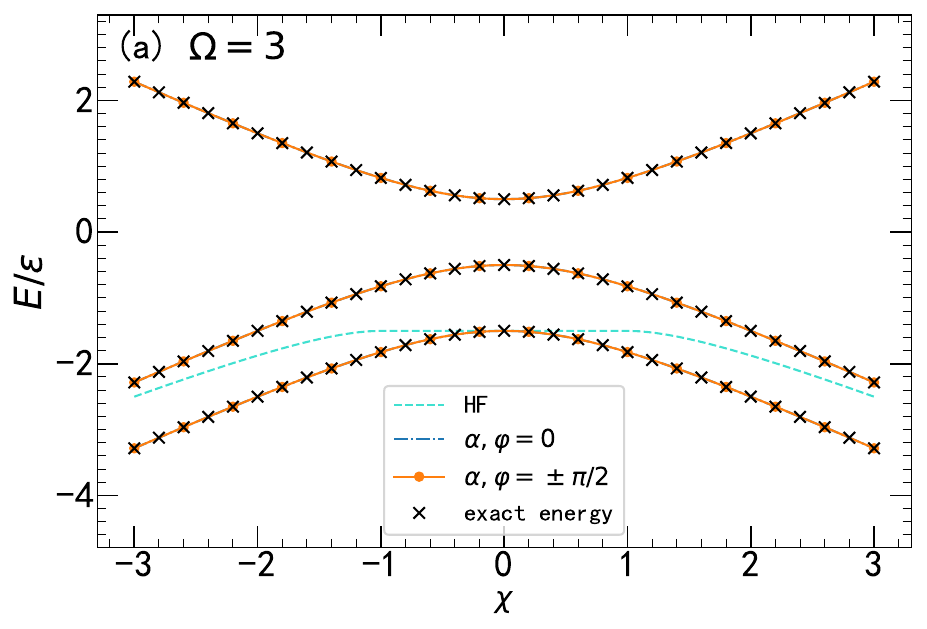} 
    \includegraphics[width=0.7\columnwidth]{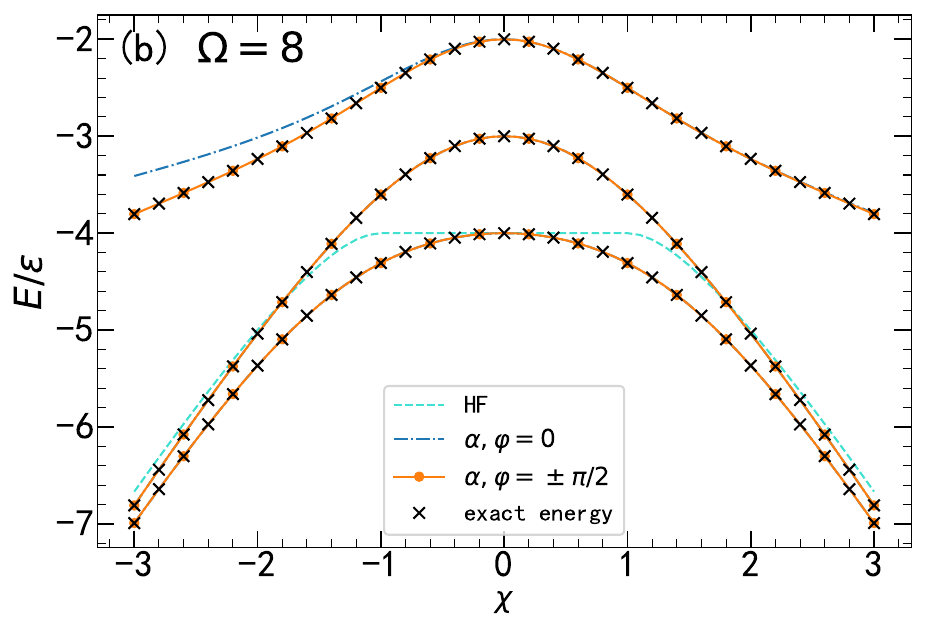} 
    \includegraphics[width=0.7\columnwidth]{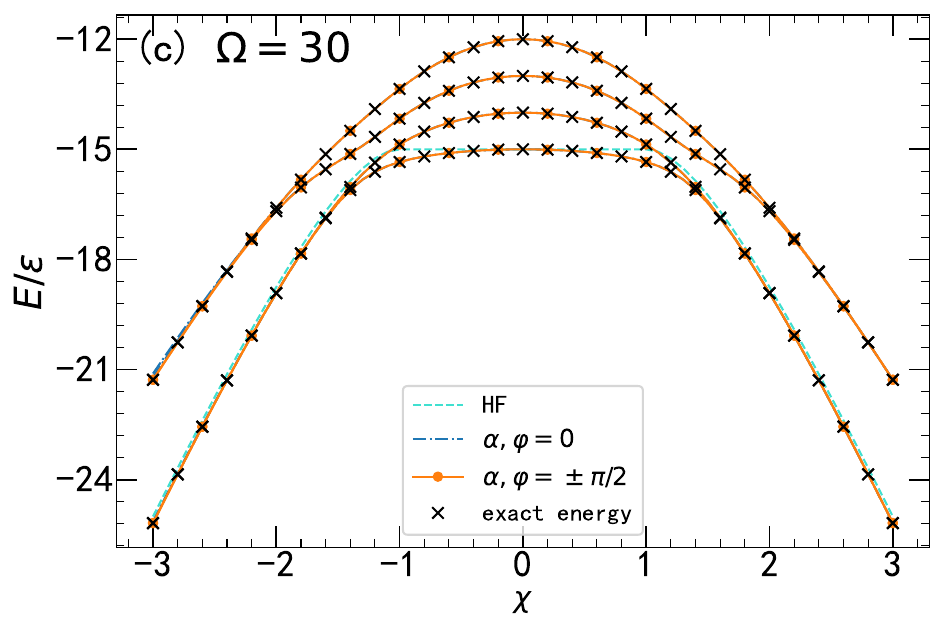} 
    \caption{(Color online) Comparison between the energies of low-lying states obtained from exact solutions, two types of GCM calculations, and HF solutions for (a) $\Omega = 3$, (b) $\Omega = 8$, and (c) $\Omega = 30$ particles, respectively. }  
    \label{fig:HF_GCM_LMG}
\end{figure}

We first examine GCM calculations for the low-lying states in the LMG model.  Fig.~\ref{fig:E_plateau_GCM} displays the convergence behaviors of the energies of the low-lying states for $\Omega=30$ particles in three types of GCM calculations against the number of natural states, where the interaction strength parameter is chosen as $\chi=-3$. In Fig.~\ref{fig:E_plateau_GCM}(c), it is demonstrated that selecting generator coordinates $\mathbf{q}(\alpha, \varphi)$ with three different values of $\varphi(0, \pm\pi/2)$ may generate spurious states. Additionally, the choice of $\varphi=0$ can also result in incorrect solutions for certain low-lying states of other Hamiltonians, see Fig.~\ref{fig:HF_GCM_LMG}. The results of HF calculation and two types of GCM calculations with the choices of $\varphi=0$ and $\varphi=\pm\pi/2$ are compared with exact energies for different $\hat H(\chi)$. It is seen that the choice of $\varphi=\pm\pi/2$ not only yields exact energies for states but also results in the fastest convergence rate, see Fig.~\ref{fig:E_plateau_GCM}(b).   Therefore, in the subsequent calculations, we choose $\varphi=\pm\pi/2$.  Moreover, it is interesting to note from Fig.~\ref{fig:HF_GCM_LMG} that the HF energy is getting closer to the ground-state energy with the increase of the number of particles from $\Omega=3$ to $\Omega=30$.

\begin{figure}
    \centering
    \includegraphics[width=0.9\columnwidth]{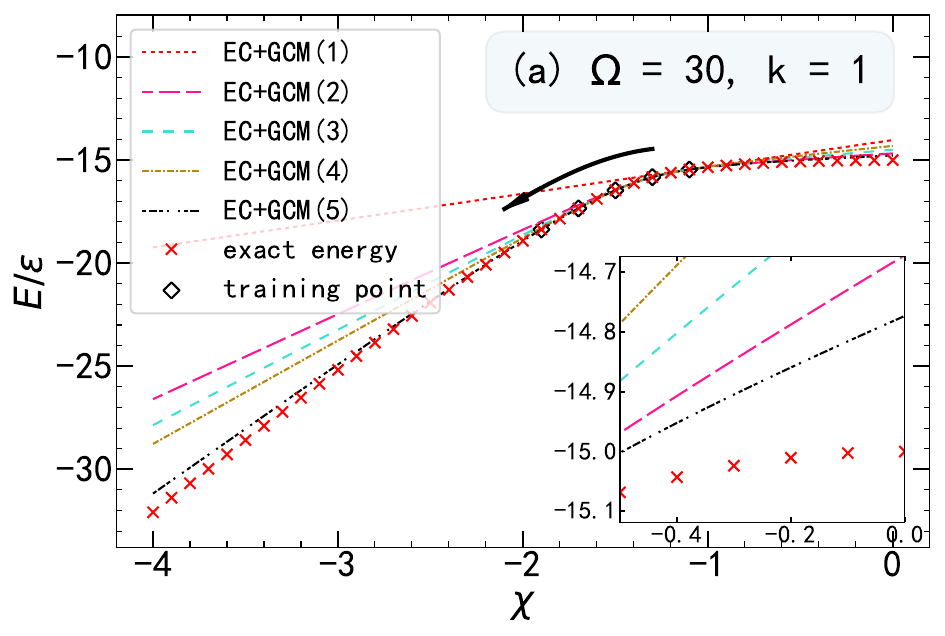}
    \includegraphics[width=0.9\columnwidth]{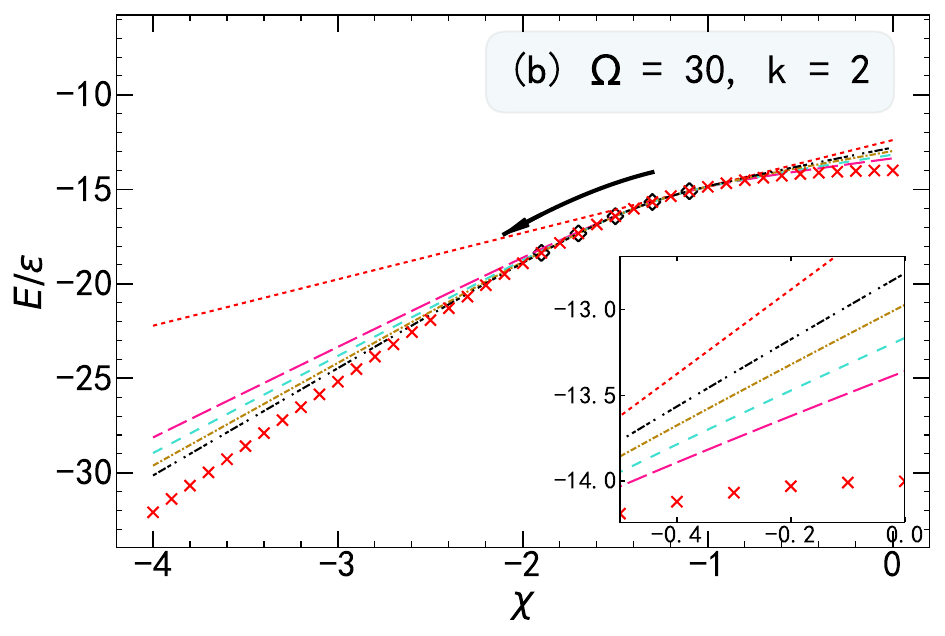}
    \includegraphics[width=0.9\columnwidth]{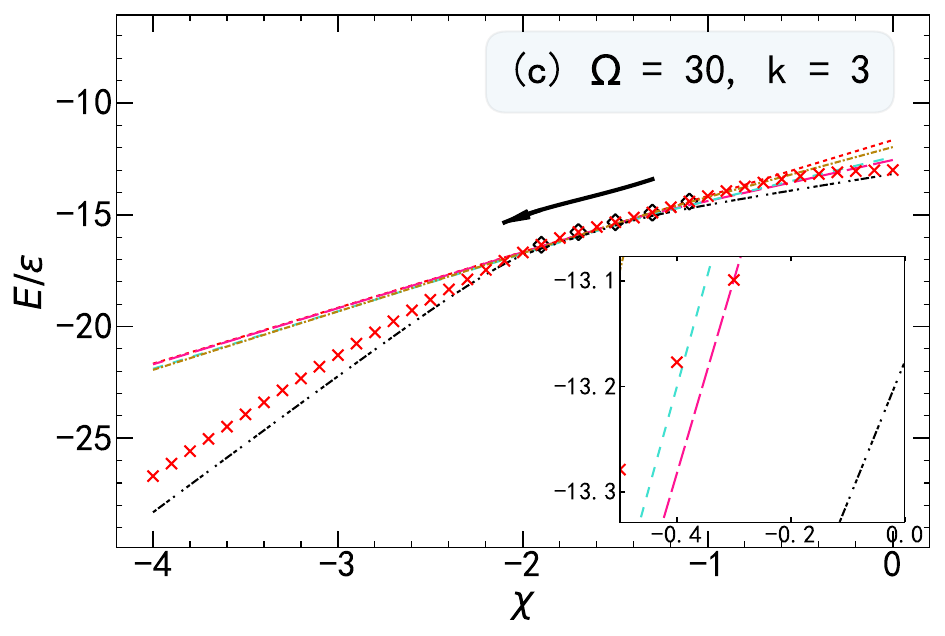}
    \caption{(Color online) Comparison between the energies of the first three states of the LMG model for $\Omega=30$ particles obtained from the exact solution and EC$_1$+GCM calculations. The results of EC$_1$+GCM calculations using different numbers (the arrow indicates the direction of increasing the training vectors from 1 to 5) of training Hamiltonians with the interaction parameter $\chi=-1.1, -1.3, -1.5, -1.7$, and $-1.9$ are provided for comparison. See the main text for details. }
    \label{fig:EC1+GCM_k3_chi5}
\end{figure}

\subsection{Results of EC$_1$+GCM calculations}

\begin{table}[] 
\tabcolsep=12pt
\caption{The logarithm of the overlap $\log_{10}\Big(\langle \Psi^{k}_{\rm EC}\ket{\Psi^{k'}_{\rm EC}}\Big)$ between the first four states of the LMG model for $\Omega=30$ particles from the EC$_1$+GCM calculations for the target Hamiltonian  with $\chi_{\odot} = -3$. See Fig.~\ref{fig:EC1+GCM_k3_chi5} for more details.}
\begin{tabular}{l|cccc}
\hline
                              & $\ket{\Psi^1_{\rm EC}}$ & $\ket{\Psi^2_{\rm EC}}$ & $\ket{\Psi^3_{\rm EC}}$ & $\ket{\Psi^4_{\rm EC}}$ \\ 
                            \hline
$\bra{\Psi^1_{\rm EC}}$ &  0      & $-11.3$            &$ -0.2$                 & $-12.0$                     \\  
$\bra{\Psi^2_{\rm EC}}$ & -      & 0       &$ -12.0 $                   &$ -1.4 $                  \\
 
$\bra{\Psi^3_{\rm EC}}$ & -        &  -               & 0      & $-13.9$                      \\  
$\bra{\Psi^4_{\rm EC}}$ & -       &  -        &  -            & 0   \\   
\hline
\end{tabular}
\label{table:overlap}
\end{table}

\begin{figure}
    \centering
    \includegraphics[width=0.9\columnwidth]{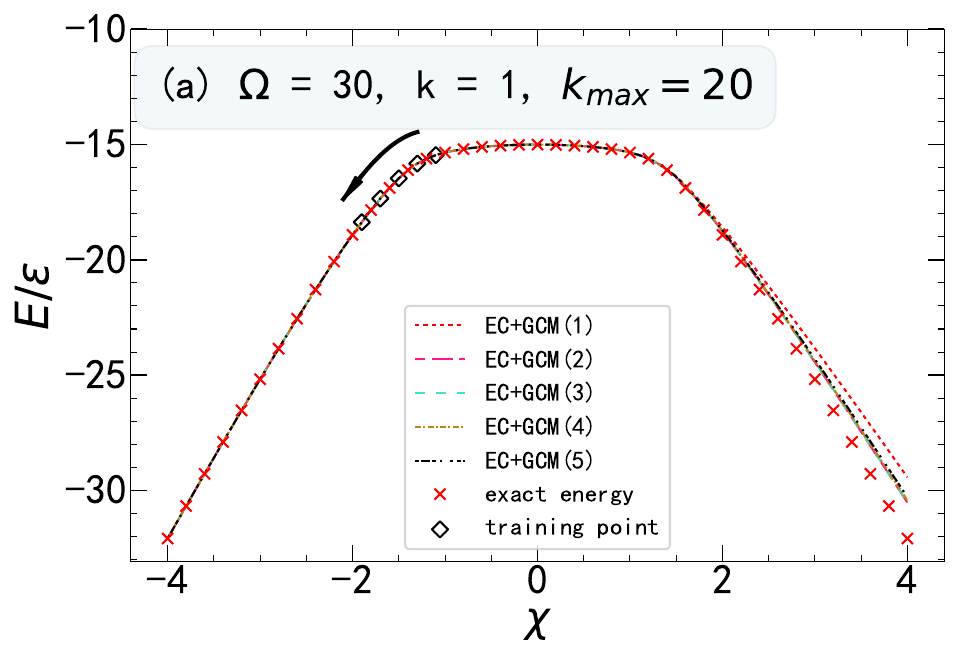}
    \includegraphics[width=0.9\columnwidth]{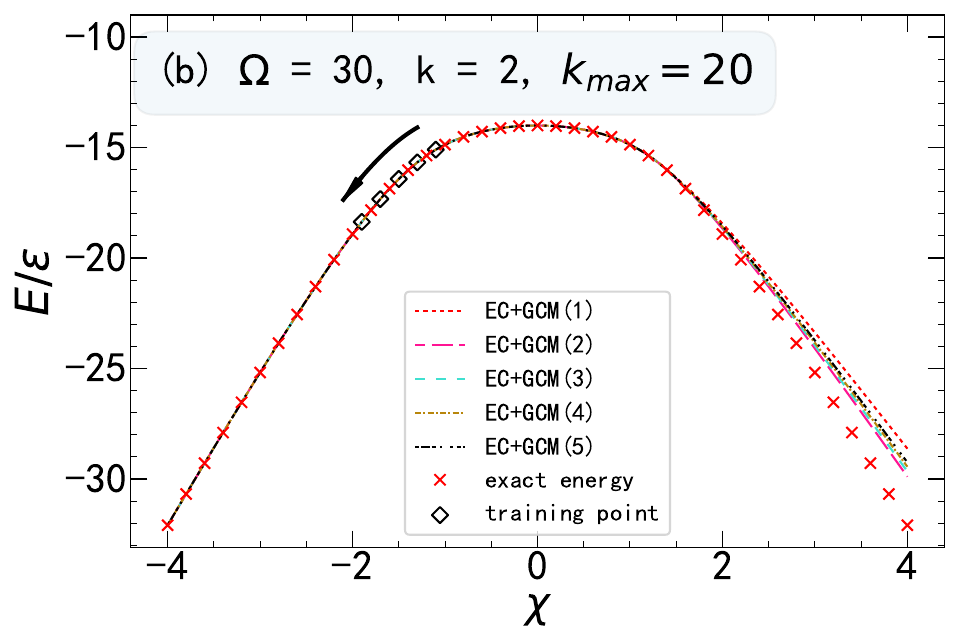}
    \includegraphics[width=0.9\columnwidth]{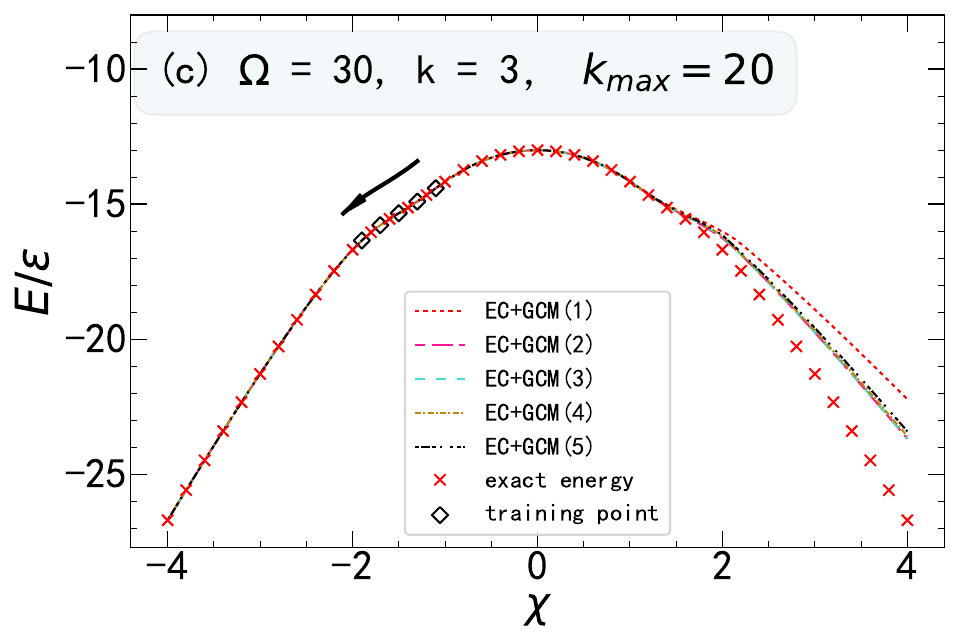} 
    \caption{(Color online) Same as Fig.~\ref{fig:EC1+GCM_k3_chi5}, except that the EC$_1$+GCM is replaced by the EC$_{\rm kmax}$+GCM, where $k_{\rm max}=20$.}
    \label{fig:ECkmax+GCM_k3_chi5}
\end{figure}

\begin{figure}
    \centering
    \includegraphics[width=0.9\columnwidth]{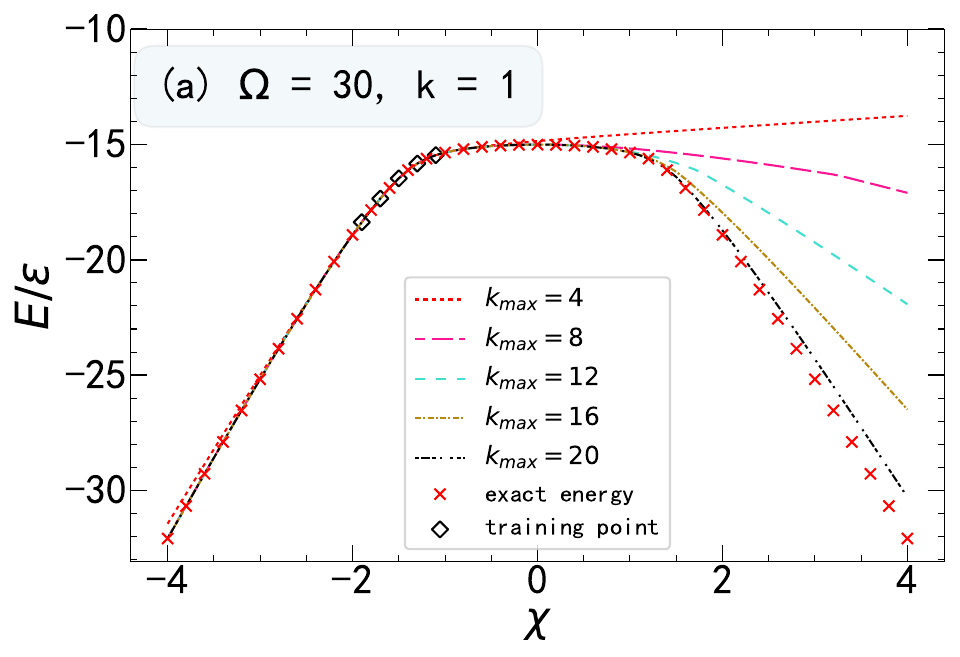}
    \includegraphics[width=0.9\columnwidth]{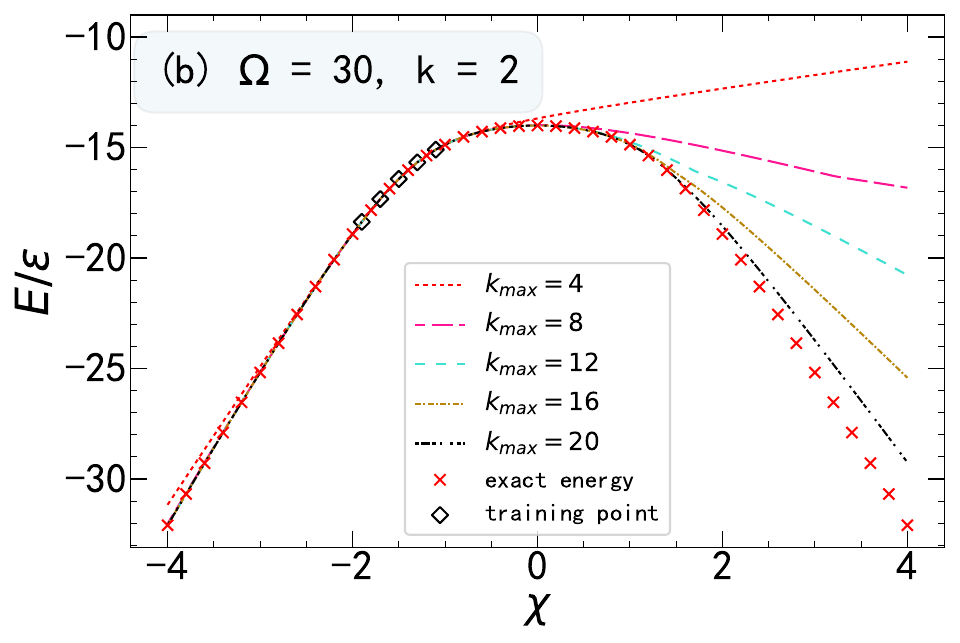}
    \includegraphics[width=0.9\columnwidth]{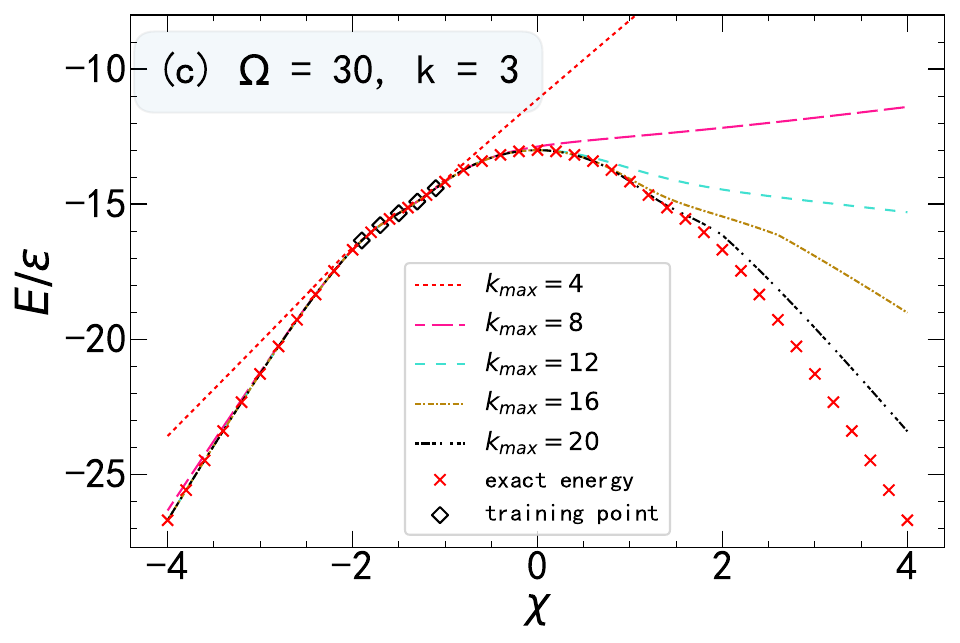} 
    \caption{(Color online) Same as Fig.~\ref{fig:ECkmax+GCM_k3_chi5}, except that different values of $k_{\rm max}$ are used in the EC$_{\rm kmax}$+GCM calculations.}
    \label{fig:converge_ECkmax+GCM_k3_chi5}
\end{figure}

In this subsection, we examine the validity of the EC$_1$+GCM calculation for the low-lying states of the LMG model with $\Omega=30$ using different Hamiltonians. Fig.~\ref{fig:HF_Energy} illustrates that changing the value of $\chi$ can simulate the system undergoing a phase transition from non-collective to collective states. Let's first consider the Hamiltonians with $\chi<-1$. The EC basis consists of the wave functions of five sampling Hamiltonians with $\chi=-1.1, -1.3, -1.5, -1.7$, and $-1.9$, respectively. The results of EC$_1$+GCM calculations for the first three states are shown in Fig.~\ref{fig:EC1+GCM_k3_chi5}. For comparison, we increase the number of sampling Hamiltonians from one to five. It is observed that increasing the number of sampling Hamiltonians does not always improve the agreement with the exact energies, especially for the excited states. A similar phenomenon is found for other values of $\chi$. Particularly, we find it challenging to reproduce the energies of the target Hamiltonians with $\chi>0$ using the wave functions of sampling Hamiltonians with $\chi<0$. We further examine the wave functions of different low-lying states from the EC$_1$+GCM calculations, which are found to be nonzero in some cases. Table~\ref{table:overlap} presents the logarithm of the overlaps $\log_{10}\Big(\langle \Psi^k_{\rm EC}\ket{\Psi^{k'}_{\rm EC}}\Big)$ between the first four states from the EC$_1$+GCM calculations for the target Hamiltonian with $\chi_{\odot} = -3$. It is shown that the overlap between the $k$-th and $k'$-th states is sizable if both $k$ and $k'$ are even or odd. Thus, one needs to ensure the orthogonality of the wave functions of different low-lying states for a given Hamiltonian in the EC method.

\subsection{Results of EC$_{\rm kmax}$+GCM calculations}

\begin{figure}
    \centering
    \includegraphics[width=\columnwidth]{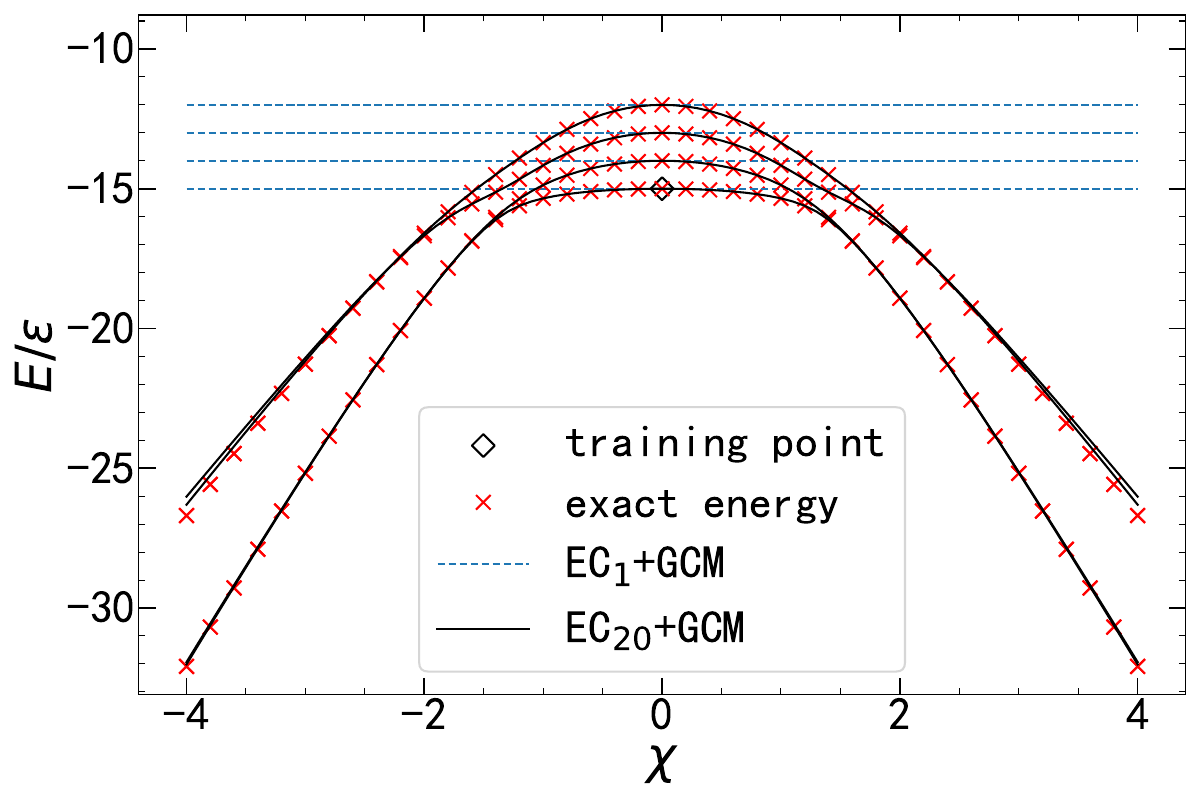}
    \caption{(Color online) 
Comparison between the energies of the first four states of the LMG model for $\Omega=30$ particles obtained from the exact solution (red crossings), EC$_1$+GCM (blue dashed curves), and EC$_{\rm kmax=20}$+GCM (black solid) calculations for different target Hamiltonians $\hat H(\chi_\odot)$. In both types of EC+GCM calculations, only one sampling Hamiltonian (indicated with an open diamond) with $\chi_t=0$ is used. }
    \label{fig:one4all_comparison}
\end{figure}

\begin{figure}
    \centering
    \includegraphics[width=0.9\columnwidth]{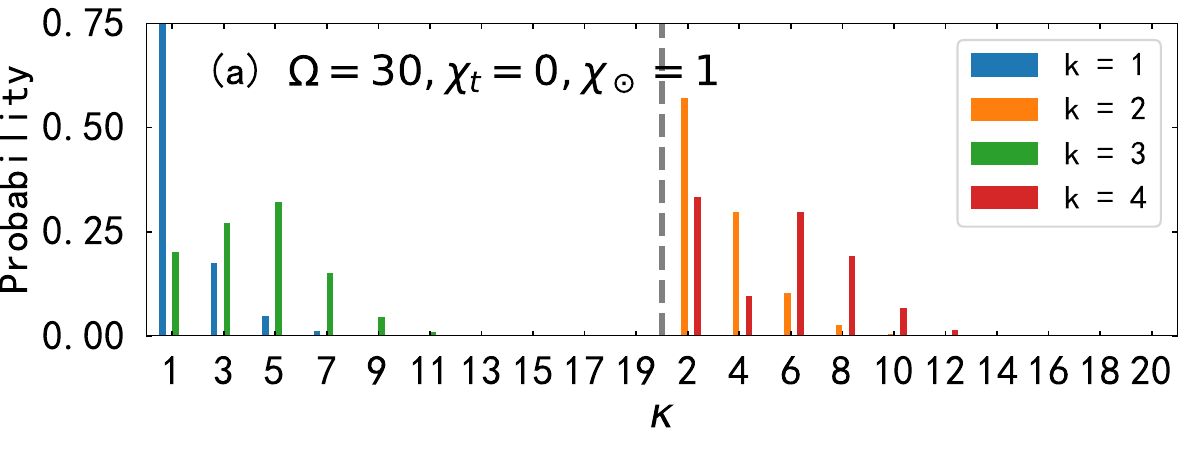}
    \includegraphics[width=0.9\columnwidth]{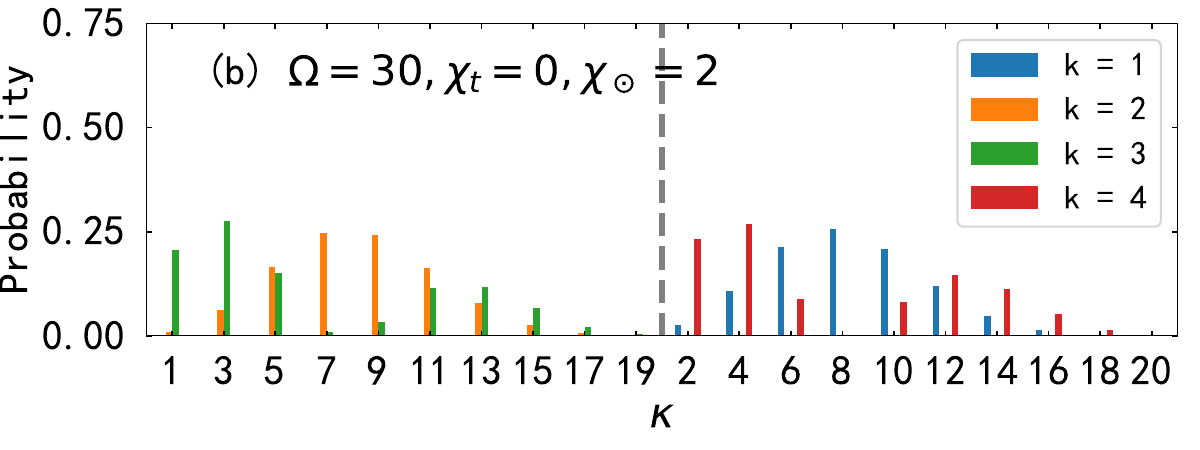}
    \includegraphics[width=0.9\columnwidth]{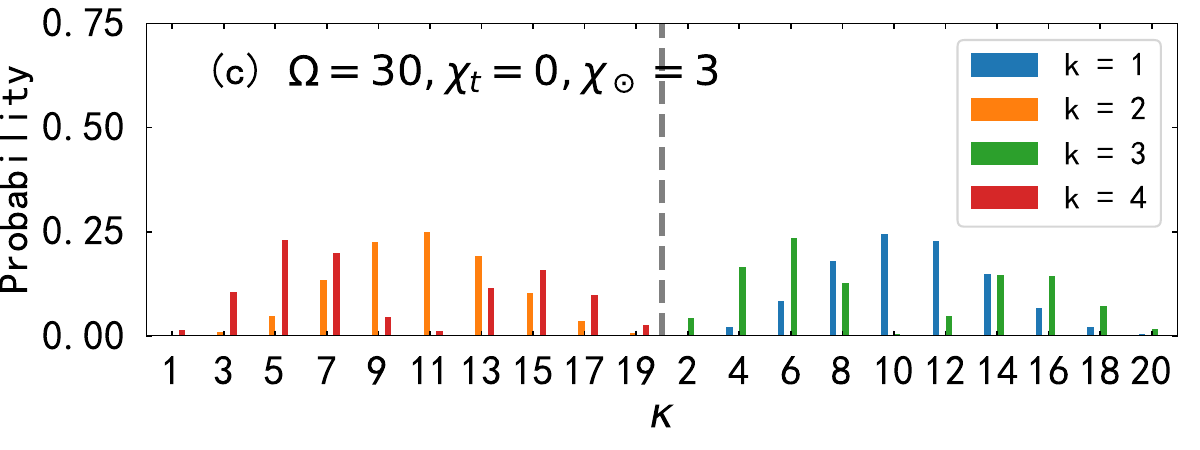}
    \caption{(Color online) Probability distribution of the first four states of the target Hamiltonian $\hat H(\chi_\odot)$ with $\chi_\odot=1, 2, 3$, respectively, over the EC basis, which correspond to the first 20 states of the sampling Hamiltonian $\hat H(\chi_t=0)$. See main text for details.}
    \label{fig:probability_dist}
\end{figure}

\begin{figure}
    \centering
    \includegraphics[width=\columnwidth]{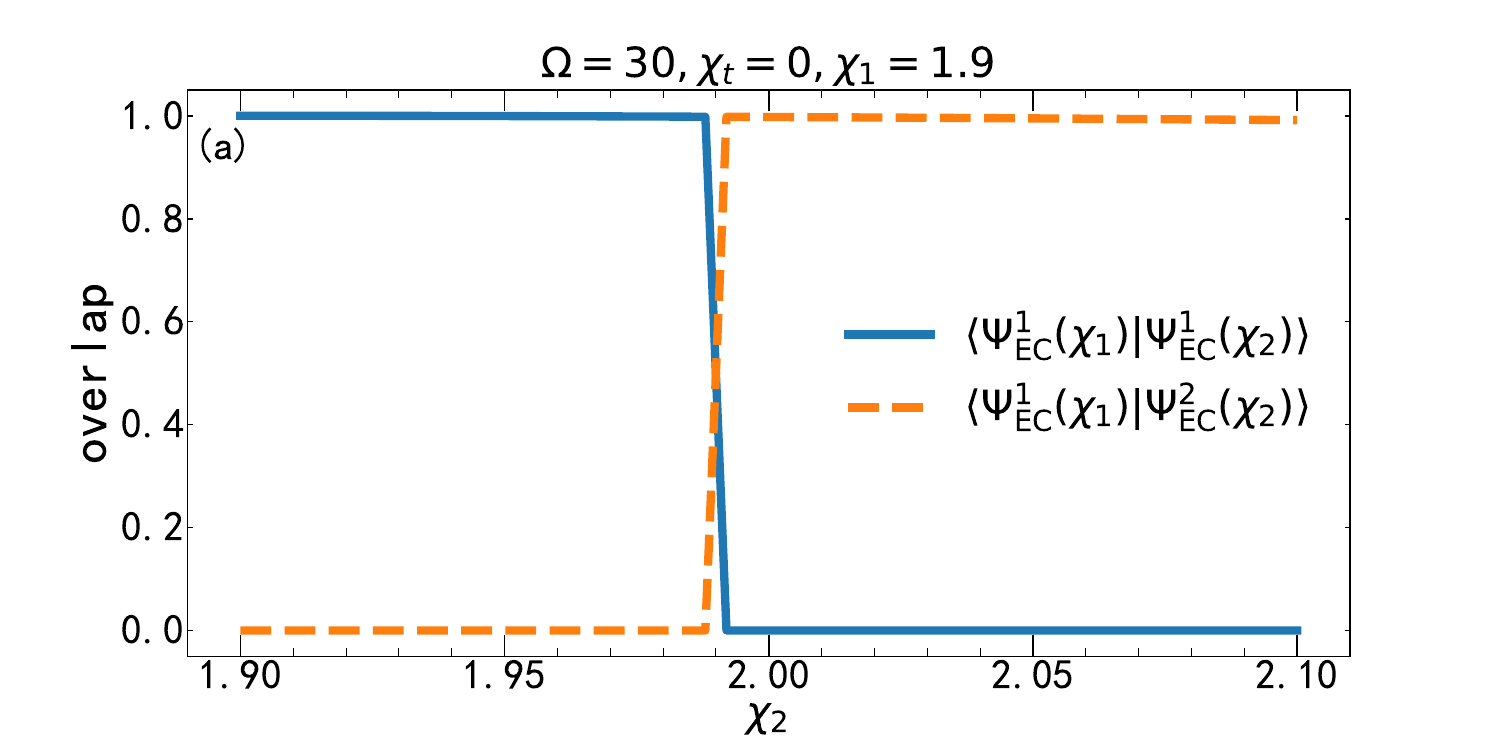}
    \includegraphics[width=\columnwidth]{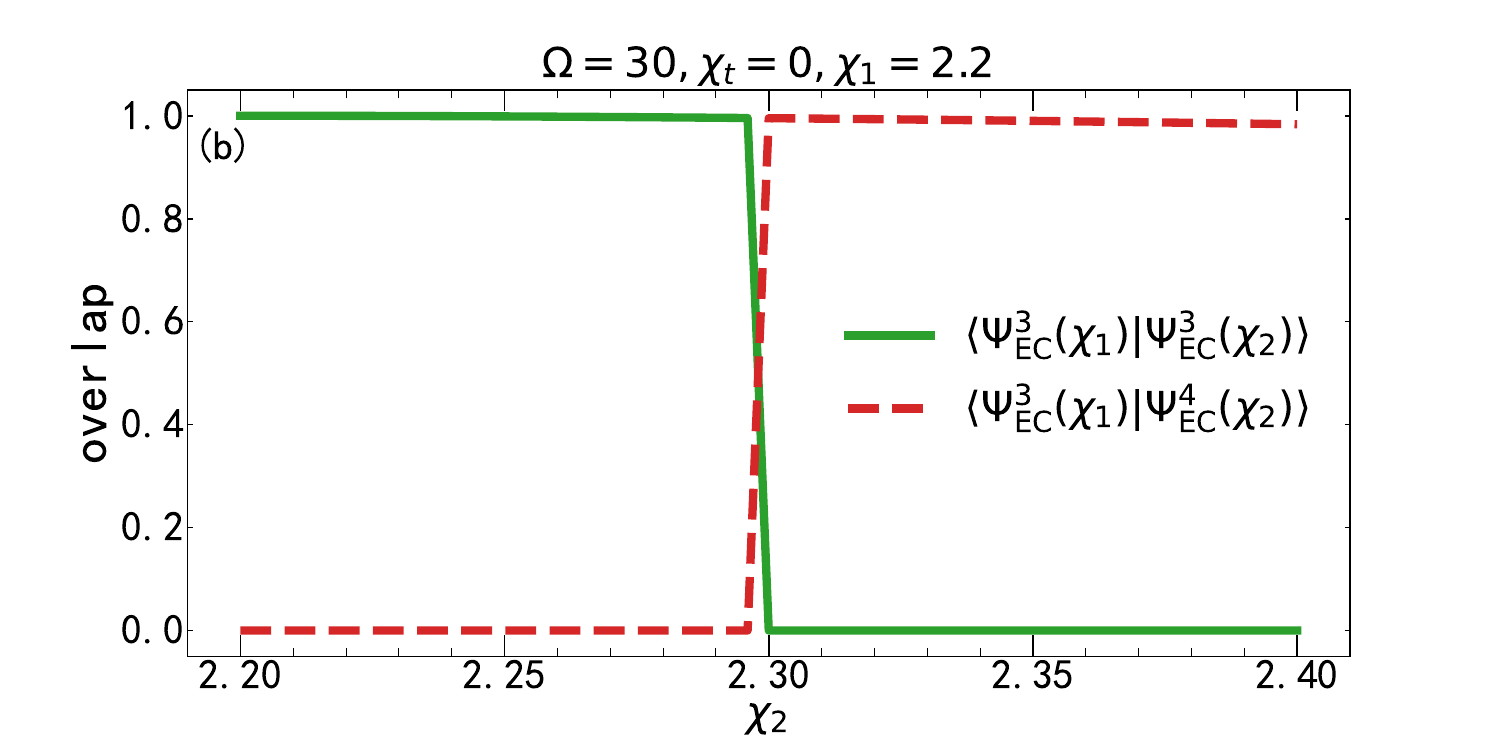}
    \caption{(color online) (a) Overlaps of the wave functions of the first two states of the target Hamiltonian $\hat H(\chi_2)$ with the first state of the Hamiltonian $\hat H(\chi_1=1.9)$, (b) and the overlaps of the wave functions of the 3rd and 4th states of the target Hamiltonian $\hat H(\chi_2)$ with the 3rd state of the Hamiltonian $\hat H(\chi_1=2.2)$, as a function of the parameter $\chi_2$. All eigenvectors are expanded based on the states of the sampling Hamiltonian $\hat H(\chi_t=0)$. See text for details.}
    \label{fig:overlap_plot}
\end{figure}

The orthogonality condition between the wave functions of different low-lying states is automatically fulfilled in the EC$_{\rm kmax}$ scheme as all the states share the same set of EC basis. Figure~\ref{fig:ECkmax+GCM_k3_chi5} presents the results of EC$_{\rm kmax}$+GCM calculations for the same systems as those in Fig.~\ref{fig:EC1+GCM_k3_chi5}. With the choice of $k_{\rm max}=20$, the exact energies of the first three states by the Hamiltonians with different values of $\chi$ are excellently reproduced in the EC$_{\rm kmax}$+GCM calculations, even in the case with only one sampling Hamiltonian ($N_t=1$). It is demonstrated in Fig.~\ref{fig:converge_ECkmax+GCM_k3_chi5} that with a sufficient number of $k_{\rm max}$, the low-lying states of the Hamiltonian with $\chi>0$ can be well represented even using the wave functions of the Hamiltonians $\hat H(\chi)$ with $\chi<0$.

Figure~\ref{fig:one4all_comparison} illustrates the performance of the extreme case in the EC$_{\rm kmax}$ scheme, i.e., with $k_{\rm max}=20$ and only one sampling Hamiltonian $\hat H(\chi_t=0)$. For comparison, the results of calculations with the EC$_1$ scheme are also plotted. It is shown that the EC$_1$+GCM(1) totally fails to reproduce the energies of states. In contrast, the EC$_{20}$+GCM(1) is able to reproduce the low-energy spectra of different Hamiltonians with $\chi\in[-4, 4]$. According to Eq.(\ref{eq:time_complexity_EC}, the time complexity of EC+GCM with only $N_t=1$ is simplified into $O\Big( N^2_q \Big)\Delta T_1$. In this case, the generalized eigenvalue problem in the EC becomes the standard eigenvalue problem, where the norm kernel in the EC is an identity matrix.  It demonstrates the super advantages of the  EC$_{\rm kmax}$  scheme in emulating GCM calculations, compared to the EC$_1$ that has been frequently employed in the literature.  In particular, we note that the EC$_{\rm kmax}$  scheme is able to reproduce the low-lying states with level crossings.

Figure~\ref{fig:probability_dist} displays the probability distribution of each EC basis in the first four states for three different target Hamiltonians with $\chi_\odot=1, 2$, and $3$, respectively. The EC basis comprises the first 20 GCM states of the Hamiltonian $\hat H(\chi_t=0)$. It is evident that as the interaction strength increases, all the first four states exhibit a broader distribution over the EC basis, highlighting the importance of including the wave functions of excited states of the sampling Hamiltonian.

Figure~\ref{fig:overlap_plot}(a) displays the overlaps of the wave functions of the first two states of the Hamiltonian $\hat H(\chi_2)$ with the first state of the Hamiltonian $\hat H(\chi_1=1.9)$ as a function of $\chi_2$. The overlaps of the wave functions of the third and fourth states of the Hamiltonian $\hat H(\chi_2)$ with the third state of the Hamiltonian $\hat H(\chi_1=2.2)$ are shown in Fig.\ref{fig:overlap_plot}(b). It is evident that there are level crossings around $\chi=1.99$ and $2.30$, respectively, which explain the exchange of the predominant components of the first two states of the target Hamiltonian with $\chi_\odot=2$ in Fig.\ref{fig:probability_dist}(b) from odd (even) to even (odd)-number indexed basis, as well as the exchange of the components of the 3rd and 4th states in Fig.~\ref{fig:probability_dist}(c). In summary, the EC$_{\rm kmax}$ scheme, even with only one sampling Hamiltonian, can reproduce the level crossings in the low-lying states of the LMG model.

\section{Summary}
\label{sec:summary}

In this study, we integrated the eigenvector continuation (EC) method into the generator coordinate method (GCM) to investigate the low-lying states of the Lipkin-Meshkov-Glick (LMG) model. We compared the results obtained using two different EC schemes. In contrast to the commonly used EC$_1$ scheme, which utilizes the wave functions of the $k$-th states from sampling Hamiltonians to expand the $k$-th states of the target Hamiltonian, the EC$_{\rm kmax}$ scheme, incorporating the wave functions of low-lying states up to the $k_{\rm max}$-th ($k_{\rm max}>k$) state of sampling Hamiltonians into the EC basis, demonstrates superior performance in terms of both efficiency and accuracy in the GCM calculations, even for states of different phases. Our findings showcase the remarkable capability of EC$_{\rm kmax}$+GCM in accurately reproducing the low-lying states of the LMG model with only a few sampling Hamiltonians. This investigation underscores the promising application of the EC method as an efficient emulator of GCM-based approaches for nuclear low-lying states, which could be utilized to quantify the statistical uncertainties of calculations for the observables of interest in future studies.

\section*{Acknowledgments} 

We thank K. Hagino, H. Hergert, D. Lee,  H. Z. Liang, C.F. Jiao, X.L. Zhang, and Y.N. Zhang  for fruitful discussions.  This work is partly supported by the National Natural Science Foundation of China (Grant Nos. 12141501 and 12275369), Guangdong Basic and Applied Basic Research Foundation (2023A1515010936) and the Fundamental Research Funds for the Central Universities, Sun Yat-sen University.
 
 \bibliographystyle{apsrev4-1}

   
%

\end{document}